\documentclass[10pt]{article}

\usepackage[margin=0.9in]{geometry}
\usepackage{amsmath, amssymb, amsfonts, mathtools, bm}
\usepackage{longtable, multirow, booktabs}
\usepackage{enumitem}
\usepackage{hyperref}
\usepackage{xcolor}
\usepackage{titlesec}
\usepackage{tikz}
\usepackage{parskip}
\usepackage{caption}
\usepackage{graphicx}
\usepackage{pdflscape}
\usepackage{adjustbox}
\usepackage{float}
\hypersetup{
    colorlinks = true,
    linkcolor = blue,
    urlcolor = blue,
    citecolor = red
}

\usepackage[object=vectorian]{pgfornament}

\definecolor{sectioncolor}{HTML}{1A237E}      
\definecolor{subsectioncolor}{HTML}{1565C0}   
\definecolor{subsubsectioncolor}{HTML}{00796B} 

\titleformat{\section}
  {\color{sectioncolor}\normalfont\Large\bfseries}
  {\thesection}{1em}{}
\titleformat{\subsection}
  {\color{subsectioncolor}\normalfont\large\bfseries}
  {\thesubsection}{1em}{}
\titleformat{\subsubsection}
  {\color{subsubsectioncolor}\normalfont\large\bfseries}
  {\thesubsubsection}{1em}{}

\renewcommand{\arraystretch}{1.2}

\title{\textbf{\LARGE Understanding Carbon Trade Dynamics:\\A European Union Emissions Trading System Perspective}}

\author{
    Avirup Chakraborty\thanks{I thank Prof. Diganta Mukherjee for his continuous guidance and advice.} \\
    \normalsize Indian Statistical Institute, Kolkata \\
    \normalsize \hypersetup{urlcolor=black}\href{mailto:avirupchakraborty358@gmail.com}{\texttt{avirupchakraborty358@gmail.com}}
}
\date{}
\begin{document}

\maketitle

\begin{abstract}
The European Union Emissions Trading System (EU ETS), the world's first and largest cap-and-trade carbon market, is a cornerstone of EU climate policy. This study provides a comprehensive empirical analysis of the EU carbon market’s efficiency, price dynamics, and structural network from 2010 to 2020. \textbf{First}, we identify significant price clustering and short-term return predictability using an AR-GARCH model, achieving around 60\% directional accuracy and a 80\% hit rate within forecasted confidence intervals. These observed patterns motivate a deeper exploration of market structure. \textbf{Second}, leveraging this insight, a weighted network analysis of inter-country transactions uncovers a concentrated market where a few registries dominate high-value flows and exert disproportionate influence. \textbf{Finally}, building upon the network findings, country-specific log-log regressions of price on traded quantity reveal heterogeneous and sometimes counter-intuitive elasticities; in several cases, positive elasticities exceed unity, indicating that trading volumes rise with prices, a deviation from conventional demand behavior that highlights potential inefficiencies driven by speculation, strategic behavior, or policy distortions. Collectively, these results point to persistent inefficiencies within the EU ETS, including partial predictability, asymmetric market power, and anomalous price-volume relationships, implying that while the system has driven decarbonization, its trading and pricing mechanisms remain imperfect.
\end{abstract}


\section{Introduction}
The \textbf{European Union Emissions Trading System (EU ETS)} is a central component of the European Union’s climate policy framework, designed to reduce \textbf{greenhouse gas (GHG) emissions} in a cost-effective and market-oriented manner. Launched in \textbf{2005}, the EU ETS was the first large-scale carbon market in the world and has since become the \textbf{largest emissions trading system globally} in terms of both value and volume. It currently applies to \textbf{energy producers}, \textbf{energy-intensive manufacturing industries}, and \textbf{intra-European aviation}, covering sectors that are collectively responsible for nearly \textbf{50\% of the EU’s total CO$_2$ emissions}. In \textbf{2019}, the total emission cap set approximately at \textbf{1.8 billion tons of CO$_2$}.

At the core of the EU ETS lies a \textbf{cap-and-trade mechanism}, which sets a maximum limit (cap) on the total amount of greenhouse gases that can be emitted by all participating entities. Within this cap, a finite number of \textbf{EU Allowances (EUAs)} are distributed,partly through \textbf{free allocation} based on historical benchmarks, and partly through \textbf{auctioning}. Each EUA grants the right to emit one ton of CO$_2$ equivalent and is \textbf{freely tradable} on the carbon market, allowing firms to buy and sell allowances according to their emission levels and abatement costs.

Compliance under the EU ETS requires installations to \textbf{monitor, report, and verify} their annual emissions, and to \textbf{surrender} a corresponding number of allowances. Firms that emit more than their allocated allowances must purchase additional EUAs from the market, while those that emit less may sell their surplus. This \textbf{market-based design} promotes emission reductions where they are most economically efficient, thereby encouraging investment in low-carbon technologies while limiting the need for heavy-handed regulatory intervention. A central challenge for the European Union Emissions Trading System (EU ETS) is the persistence of \textbf{inefficiencies in market functioning}, despite its success as the flagship cap-and-trade mechanism for reducing greenhouse gas emissions. Market efficiency is critical for ensuring that allowance prices reflect underlying fundamentals rather than speculative or strategic behavior. Inefficiencies, however, may arise from volatility clustering, market concentration, or anomalous price-volume relationships, all of which can distort the intended cost-effective allocation of emissions reductions.

Previous \textbf{literature} has shown that carbon markets, particularly the EU ETS, often display persistent \textbf{inefficiencies} and price distortions, largely arising from liquidity constraints and market design flaws \cite{inefficiency2, inefficiency3}. Recent studies confirm that these distortions remain structural rather than temporary \cite{inefficiency}. Parallel research highlights that \textbf{liquidity} continues to shape price formation and return behavior in both EU and Chinese carbon markets \cite{zhang2022liquidity, kim2021liquidity}. Moreover, \cite{zhikai2024volatility} and \cite{zhao2024forecasting} provide evidence that carbon price movements exhibit \textbf{return predictability}, influenced by energy market spillovers and nonlinear dynamics. To facilitate such investigations, \cite{database} offer a comprehensive \textbf{international transaction database} supporting cross-country analyses of trading patterns and market efficiency.

Building on this foundation, our analysis provides new empirical evidence on the nature and extent of these inefficiencies from 2010 to 2020. This study is structured as follows: \textbf{First}, we begin with a detailed \textbf{data and market overview}, examining \textbf{transaction patterns and price dynamics} to establish the fundamental characteristics of the market and to assess the scope for \textbf{return predictability}. The observed variations in trading activity across registries and over time suggest that market behavior may not be uniform, motivating a deeper exploration of its underlying structure. \textbf{Second}, to capture this heterogeneity and the potential interdependence among participants, we conduct a \textbf{network analysis} of transaction values across registries. This enables us to uncover patterns of trading concentration and identify dominant countries whose interactions shape overall market coordination. \textbf{Finally}, building on these network insights, we focus on the top three trading nations to estimate \textbf{price elasticities}, investigating how transaction volumes respond to price changes and whether such relationships reveal inefficiencies within the system.

\section{Data and Market Overview}
\subsection{EUA Transaction Data}

The data for this study are collected from the \textbf{European Union Transaction Log} (EUTL) \cite{database}, the official registry that tracks transactions and compliance under the EU ETS. The EUTL functions as the central reporting and monitoring framework, enabling the European Commission to make key information on regulated entities’ compliance status, market participation, and allowance transfers publicly available. This data spans February 2005 to April 2020, with transaction-level records accessed through the official portal \cite{EUETS.INFO}, which provides granular emissions and compliance data at the installation level. The dataset covers three primary domains: installations, transactions, and accounts.

\textbf{Installations} correspond to regulated facilities, such as power plants, industrial plants, and aviation operators where verified emissions are reported and allowances must be surrendered to ensure compliance. \textbf{Transactions} record the transfer of allowances between participants and are tied to the account in which they occur. These transfers include internal reallocations between installations and administrative accounts, allocations through free distribution or auction mechanisms, and surrenders undertaken to meet compliance obligations. \textbf{Accounts}, in turn, are categorized into three types. \textit{Operator Holding Accounts} (OHAs), linked to installations, are used by regulated entities to receive, hold, and surrender allowances. \textit{Administrative Accounts} are managed by competent authorities to issue, allocate, or cancel allowances as part of system operations. Finally, \textit{Person Holding Accounts} (PHAs), maintained by intermediaries, financial institutions, and non-governmental organizations, are employed primarily for trading purposes but are not directly used for compliance.

The dataset records transactions of \textbf{European Union Allowances} (EUAs), the primary compliance instrument under the EU ETS. Established in 2005, the system has progressed through four phases: \textbf{Phase I} (2005–2007) as a pilot defining trading rules; \textbf{Phase II} (2008–2012) expanding scope and allocation; \textbf{Phase III} (2013–2020) introducing harmonized regulations and stricter compliance; and the ongoing \textbf{Phase IV} (2021–2030). Our study examines EUA transfers from \textbf{January 5, 2010}, to \textbf{April 30, 2020}, covering the end of Phase II and the entire Phase III of the EU ETS.

\begin{figure}[H]
    \centering
    \includegraphics[width=1\linewidth]{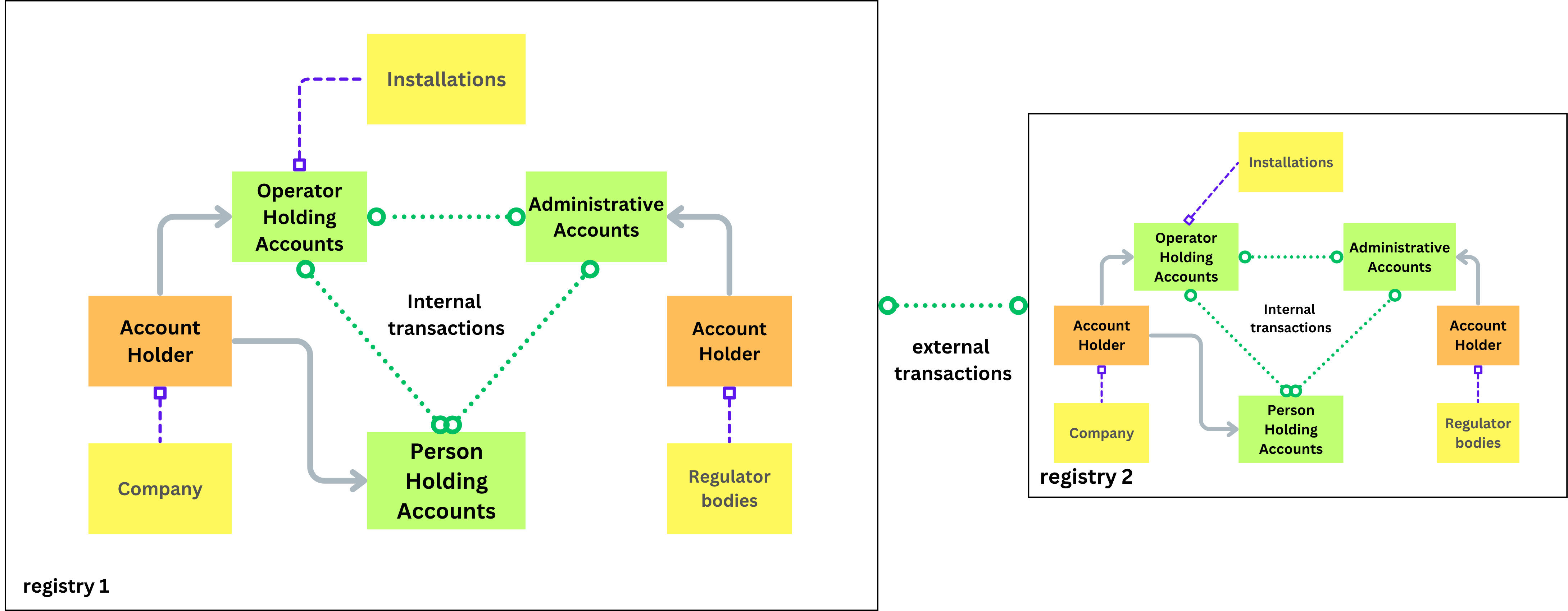}
    \caption{ }
    \label{fig:transaction}
\end{figure}

\begin{figure}[H]
    \centering
    \includegraphics[width=.9\linewidth]{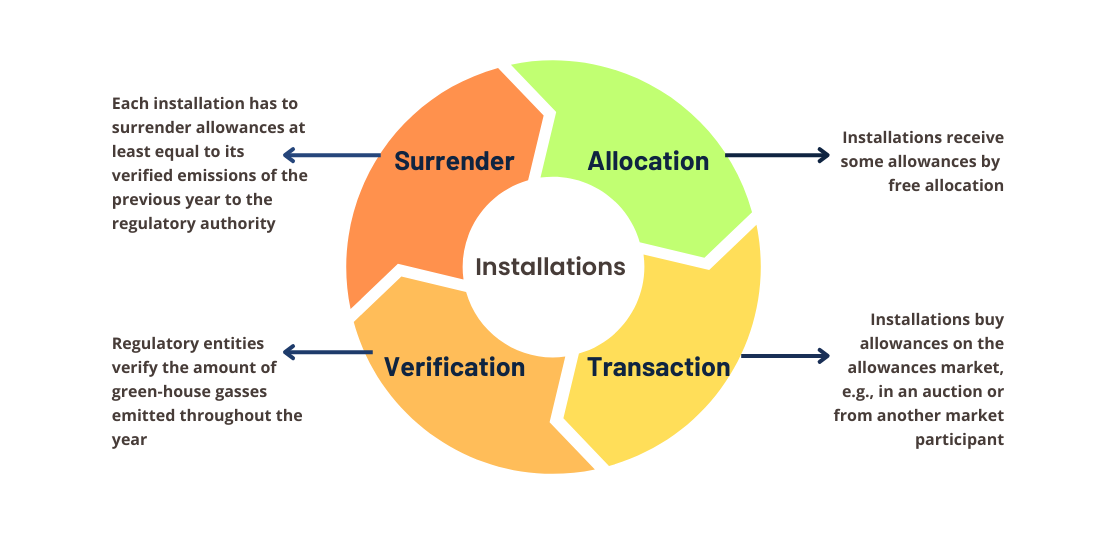}
    \caption{ }
    \label{fig:compliance}
\end{figure}

Figure~\ref{fig:transaction} illustrates the functioning of the EU ETS, highlighting the three types of accounts and their respective holders. These accounts are connected to various entities such as installations (OHA), administrative authorities (regulatory bodies), and third-party companies (PHA) that participate in market transactions to achieve compliance. The figure also distinguishes between internal and external transactions, clarifying how they differ across registries (countries).
Figure~\ref{fig:compliance} illustrates the cyclical procedure for maintaining compliance within each financial year. The process begins with the initial allocation or auctioning of emission allowances to installations, with the quantities determined by regulatory authorities. This is followed by the transaction phase, during which installations engage in market trading to either generate profit or ensure compliance. 

\begin{figure}[H]
    \centering \includegraphics[width=1\linewidth]{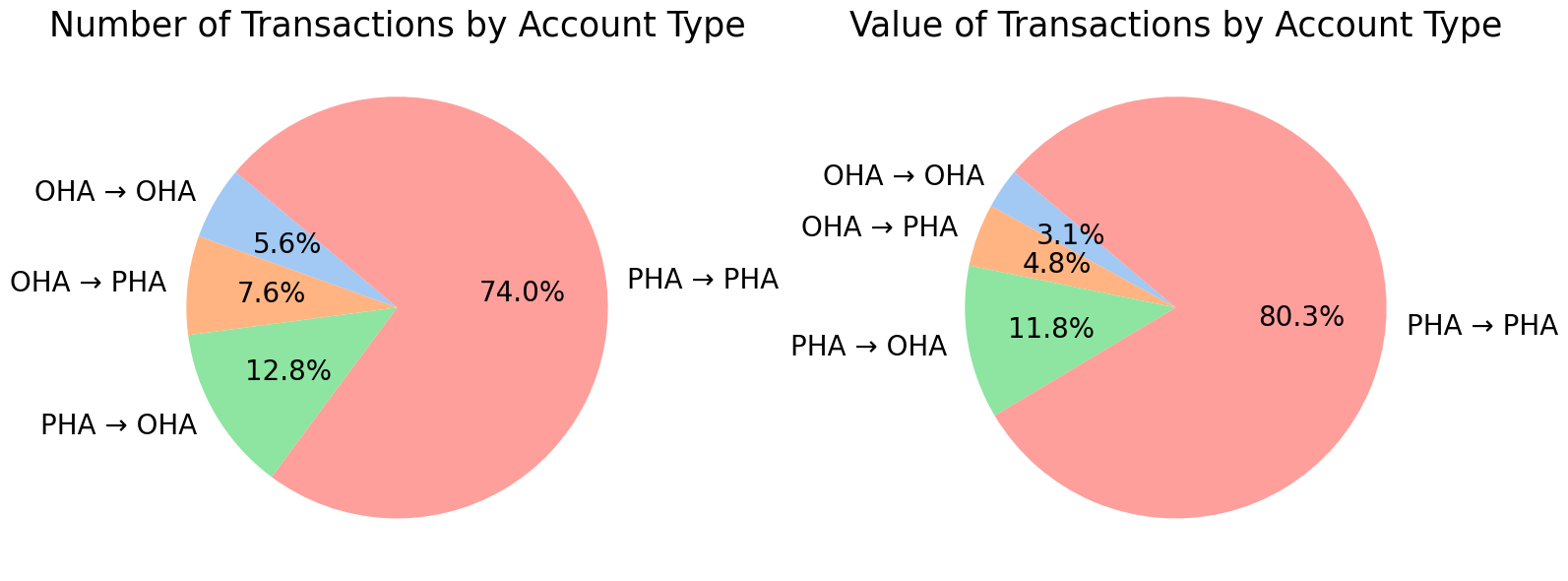}
    \caption{}
    \label{fig:descriptive1}
\end{figure}

\begin{figure}[H]
    \centering \includegraphics[width=1\linewidth]{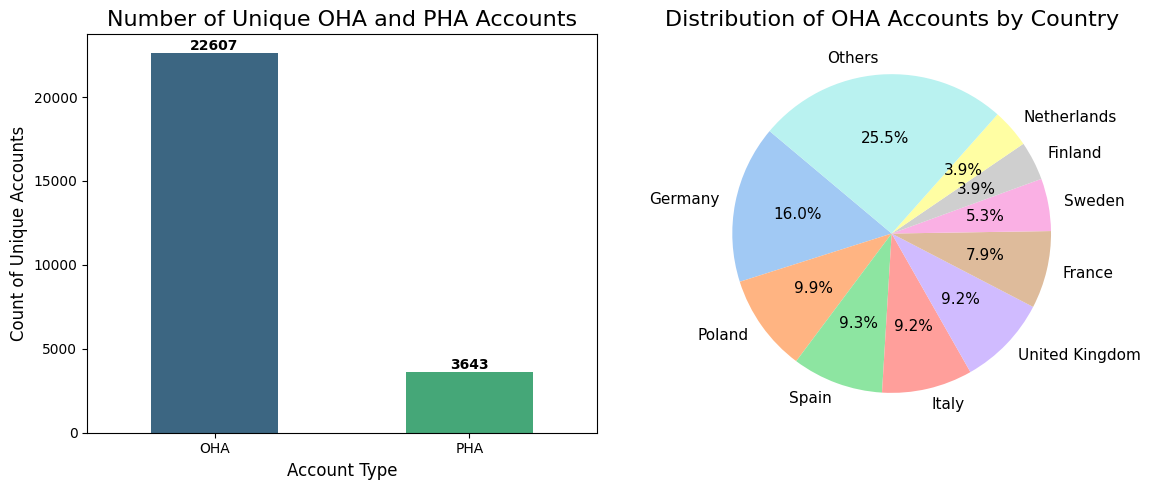}
    \caption{}
    \label{fig:descriptive2}
\end{figure}

The final two phases occur at the end of each financial year, managed by the regulatory bodies,these involve verifying the actual emissions and requiring installations to surrender the corresponding number of allowances. In the transaction data, the compliance rate is observed to be nearly 100\%, indicating effective enforcement of the system. Since administrative transactions do not effect price dynamics, our analysis focuses exclusively on transfers between OHAs and PHAs. Figure~\ref{fig:descriptive1} illustrates the proportions of transaction counts and values (in EUR) across these account types during the selected period.
Although Operator Holding Accounts (OHAs) are more numerous in the market (Figure~\ref{fig:descriptive2}, left), Person Holding Accounts (PHAs) dominate trading activity in both volume and value (Figure~\ref{fig:descriptive1}). This prominence suggests that PHAs, despite being fewer in number, play a more influential role in price formation and market dynamics. The distribution of these numerous OHA accounts by country is further detailed in Figure~\ref{fig:descriptive2} (right).

The above descriptive analysis provides a comprehensive overview of market participation and transaction dynamics. From this, we infer several key observations , notably, the limited participation of OHA accounts, which may be attributed to the pre-allocation of allowances. In Figure~\ref{fig:descriptive2} (left), the number of unique OHA and PHA accounts corresponds to the distinct accounts engaged in OHA–PHA transactions within the dataset. Approximately 80\% of all OHA accounts participated in these transactions. Thus, although OHA accounts exhibit relatively steady participation, their transaction volumes and values remain low, indicating a degree of inefficiency in market engagement. In the following section, we will now explore the dynamics of price movements in greater depth.

\subsection{EUA Price Data}

The allowance price data were collected from the \textbf{International Carbon Action Partnership} (ICAP), available at \cite{ICAP}. The dataset distinguishes between two types of markets. \textbf{The primary market} represents the initial issuance of allowances, typically through government auctions. In this market, spot prices are recorded weekly, with observations available every Tuesday. \textbf{The secondary market}, by contrast, reflects trading activity among companies, intermediaries, and other entities after allowances have been distributed in the primary market. For this market, spot price data are reported for all weekdays, excluding weekends. 

\begin{figure}[H]
    \centering \includegraphics[width=0.82\linewidth]{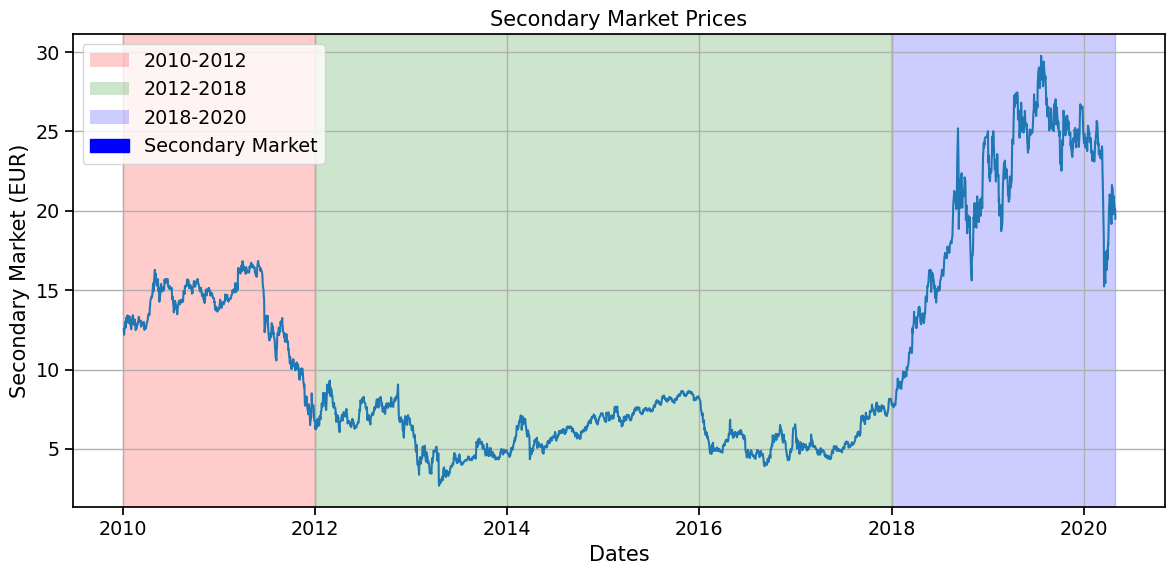}
    \caption{ }
    \label{fig:price}
\end{figure}

The backbone of the EU Emissions Trading System (EU ETS) lies in its \textbf{strategic use of market dynamics} to create an \textbf{arbitrage-free pricing mechanism}, enabling participants to \textbf{adjust and reduce carbon emissions efficiently}. For the analysis, we construct a weekly-aggregated price series by taking the mean of the secondary market spot prices within each calendar week. This approach smooths out short-term fluctuations that daily data often exhibit and provides a more stable trend that classical econometric methods can effectively capture. Figure~\ref{fig:price} presents the resulting weekly series (in Euros), partitioned into three distinct segments with different mean levels. These segments will later aid in identifying price clusters and examining the evolution of the market across phases.

The observed price behavior can be largely attributed to changes in EU ETS policy mechanisms over time. According to the European Commission, the \textbf{surplus of allowances} in the EU ETS exceeded 2.1 billion in 2013 \cite{allowance surplus}. This substantial oversupply reduced scarcity in the market, weakened demand pressures, and consequently led to a prolonged period of low EUA prices. To address this structural imbalance, the \textbf{Market Stability Reserve} (MSR) was agreed upon in 2015 as a long-term mechanism to correct the surplus of allowances in the EU carbon market. Designed to automatically adjust the supply of allowances and enhance the market’s resilience to future shocks, the MSR was formally established in 2018 and became operational in 2019 \cite{MSR}. The implementation of the MSR effectively tightened the available supply of allowances, restoring market confidence and resulting in a notable upward shift in EUA prices from 2018 onward.

Market efficiency is inherently related to the predictability of asset returns \cite{Market efficiency}. 
To examine short-term return predictability, we employ the Box-Jenkins framework to model and forecast returns. 
Specifically, the \textbf{weekly log return} is defined as  
\[
r_t = \ln\left(\frac{p_t}{p_{t-1}}\right),
\]
where \(p_t\) and \(p_{t-1}\) denote the prices at time \(t\) and \(t-1\), respectively.

\begin{figure}[h]
    \centering
    \includegraphics[width=0.72\linewidth]{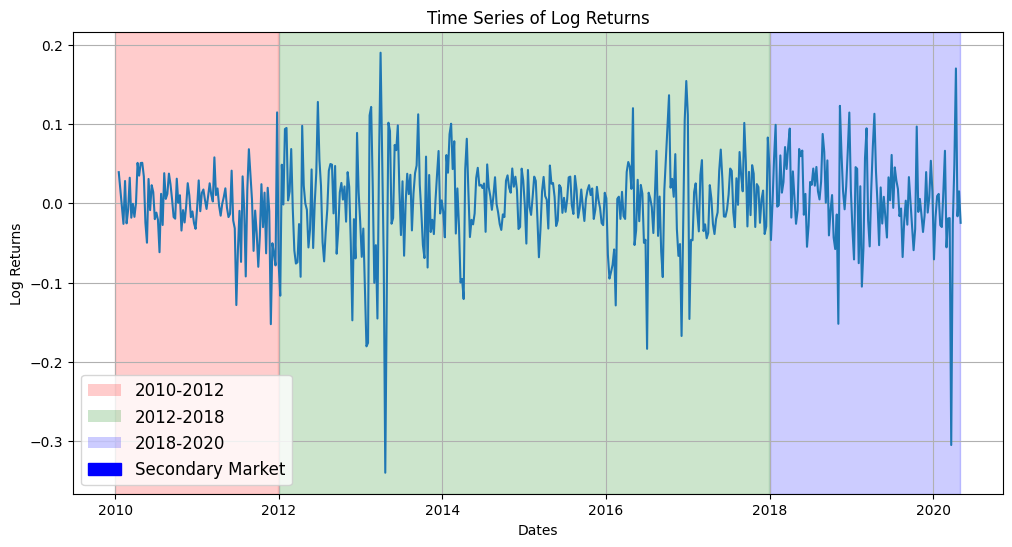}
    \caption{ }
    \label{fig:returns}
\end{figure}

Figure~\ref{fig:returns} shows the resulting return series partitioned into different periods. Since we will model the return now on, we must know about the stationarity of this series. Observing the time series it is clear that the returns are mean reversion but some heteroskedasticity is also visible. So to test the mean stationarity we will use the Augmented Dickey Fuller test and to check for time-varying variance Engle’s ARCH Test, shown in Table~\ref{tab:test}.

\begin{table}[H]
\centering
\caption{Results of Stationarity and Heteroskedasticity Tests Across Periods}
\label{tab:test}
\begin{tabular}{|l|l|c|c|c|}
\hline
\textbf{Period} & \textbf{Test} & \textbf{Test Statistic} & \textbf{p-value} & \textbf{Conclusion} \\
\hline
\multirow{2}{*}{2010--2012} 
& ADF & -5.13 & $1.20 \times 10^{-5}$ & Stationary in mean (reject $H_0$) \\
& ARCH & 24.07 & 0.0199 & No ARCH effects (fail to reject $H_0$) \\
\hline
\multirow{2}{*}{2012--2018} 
& ADF & -7.79 & $8.19 \times 10^{-12}$ & Stationary in mean (reject $H_0$) \\
& ARCH & 39.83 & $7.66 \times 10^{-5}$ & ARCH effects present (reject $H_0$) \\
\hline
\multirow{2}{*}{2018--2020} 
& ADF & -7.11 & $4.04 \times 10^{-10}$ & Stationary in mean (reject $H_0$) \\
& ARCH & 10.01 & 0.6150 & No ARCH effects (fail to reject $H_0$) \\
\hline
\end{tabular}
\end{table}

\begin{figure}[h]
    \centering
    \includegraphics[width=1\linewidth]{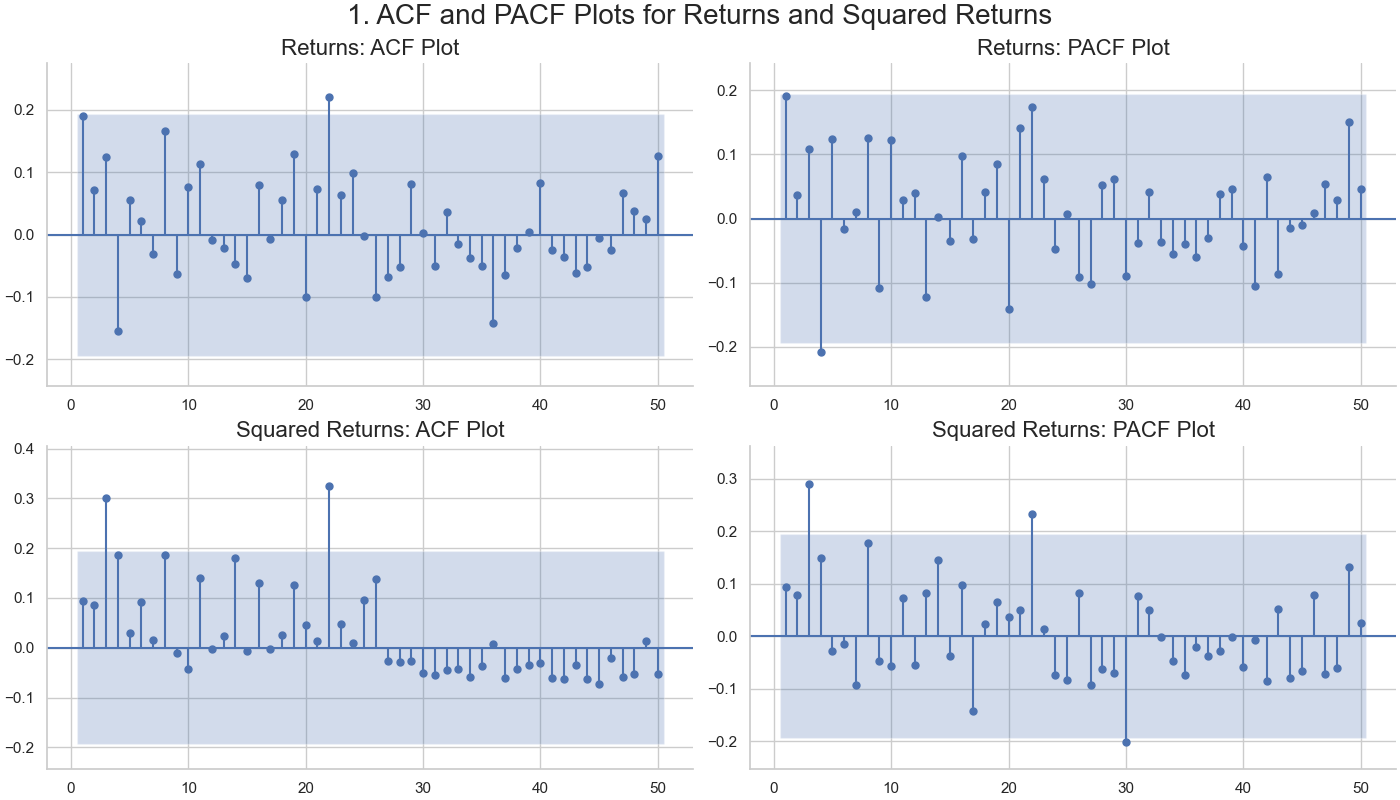}
    \caption{ }
    \label{fig:acf-pacf1}
\end{figure}

\begin{figure}[h]
    \centering
    \includegraphics[width=1\linewidth]{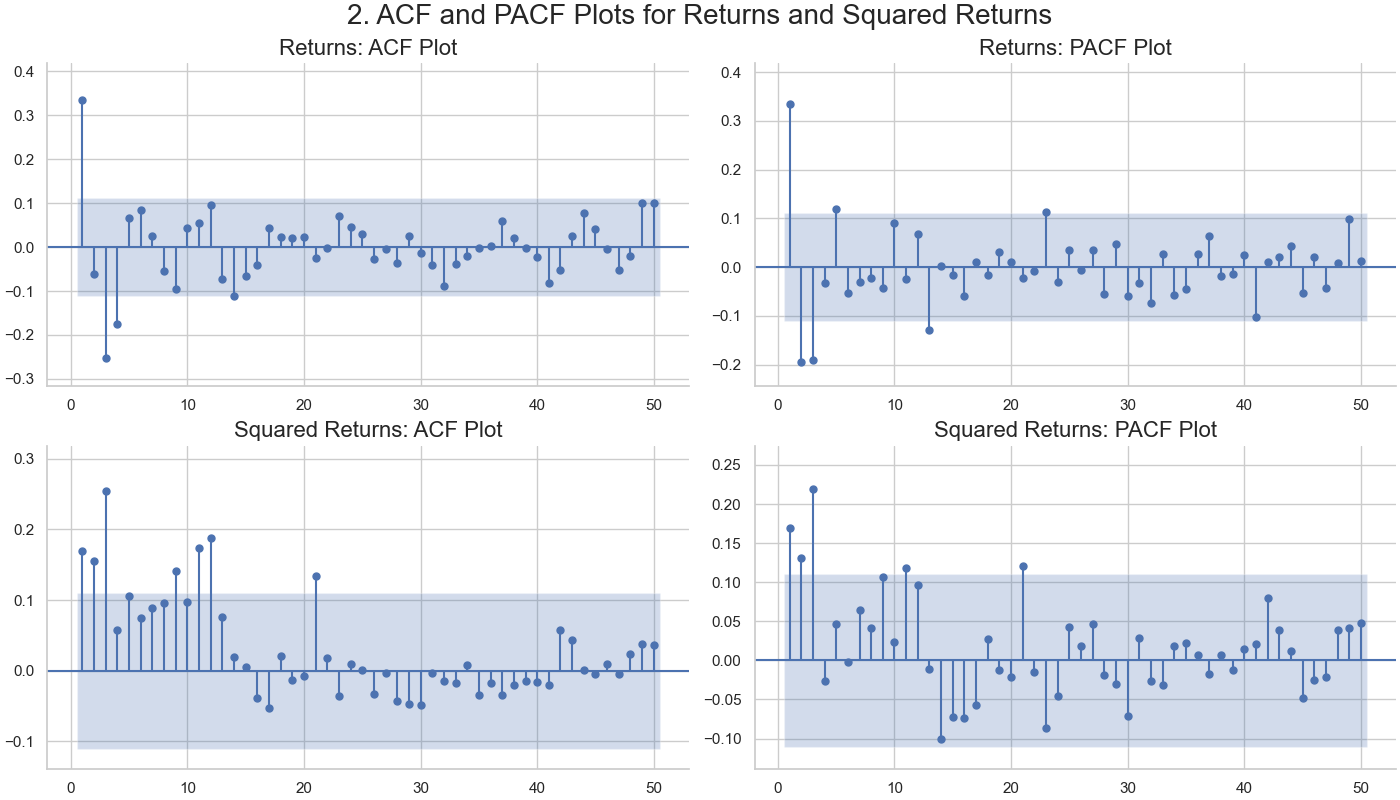}
    \caption{ }
    \label{fig:acf-pacf2}
\end{figure}

\begin{figure}[h]
    \centering
    \includegraphics[width=1\linewidth]{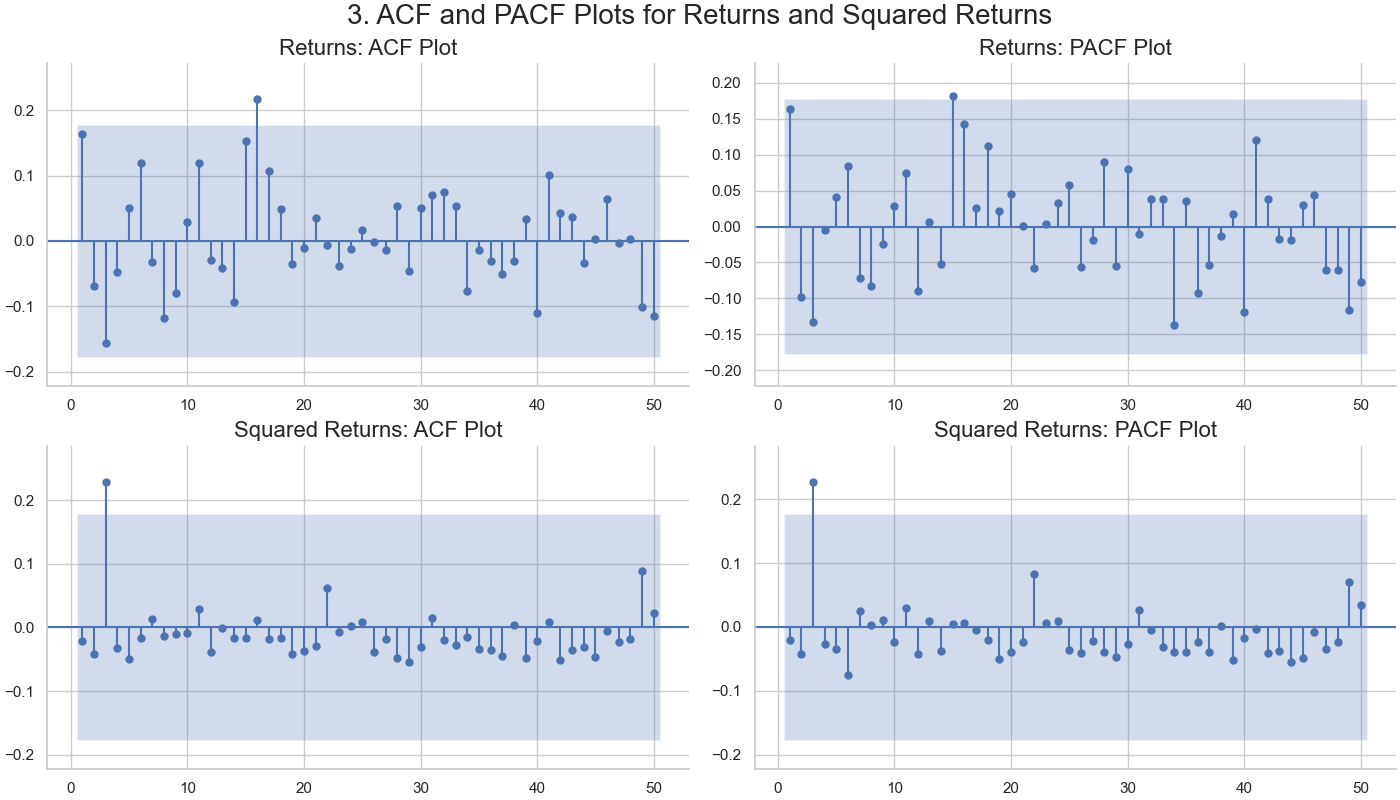}
    \caption{ }
    \label{fig:acf-pacf3}
\end{figure}

The return series is stationary in mean but exhibits time varying variance across periods. The ACF decays rapidly to zero in all cases (Figures~\ref{fig:acf-pacf1} to \ref{fig:acf-pacf3}), confirming mean stationarity, while the PACF for the second period shows significant spikes up to lag 3, indicating short run dependence and motivating an AR(3) specification for the conditional mean. Furthermore, the ACF and PACF of squared returns in the second period display strong and persistent autocorrelation, characteristic of volatility clustering, thereby providing evidence of conditional heteroskedasticity consistent with the Engle test.

\begin{figure}[H]
    \centering    \includegraphics[width=0.90\linewidth]{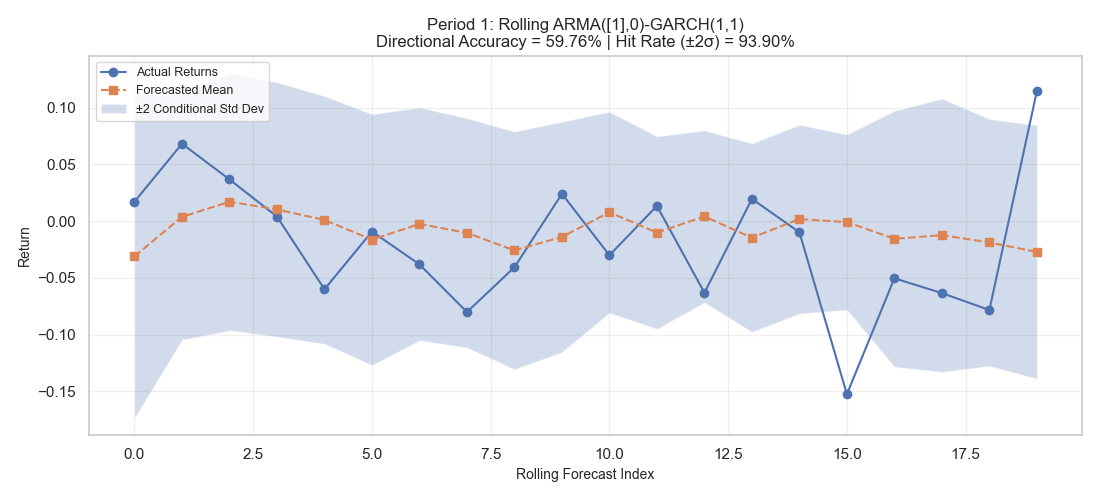}
    \caption{ }
    \label{fig:forecast1}
\end{figure}

\begin{figure}[H]
    \centering    \includegraphics[width=0.90\linewidth]{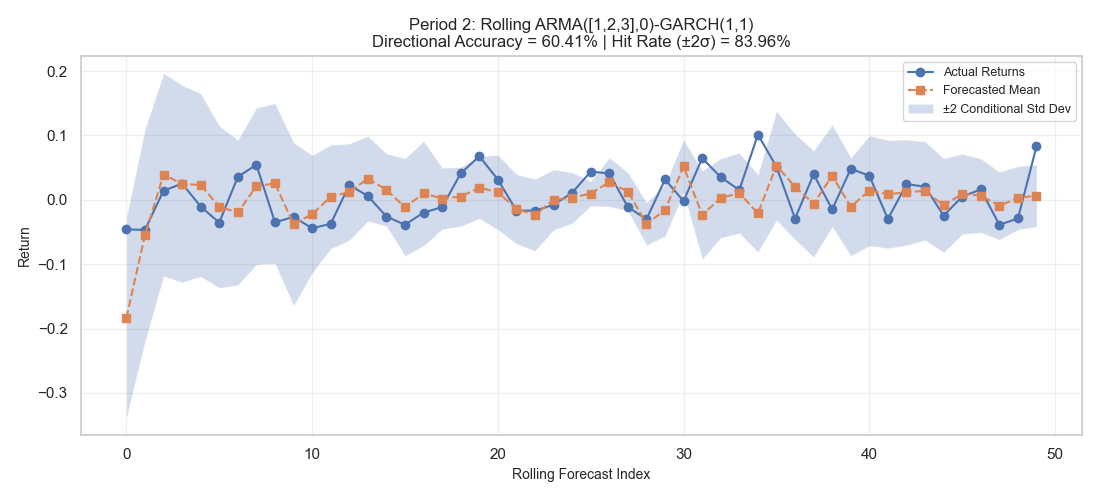}
    \caption{ }
    \label{fig:forecast2}
\end{figure}

To model these features, we implement a rolling forecasting framework with a two stage specification: period specific AR models for the conditional mean and a common GARCH(1,1) model for the conditional variance. Using a 104 week rolling window, one step ahead forecasts are generated, with residuals from the mean equation feeding into the GARCH model to obtain variance forecasts. Evaluation results (Table~\ref{tab:evaluation}) indicate that the model captures volatility reasonably well across periods, with hit rates of 93.90\%, 83.96\%, and 93.14\% for the three subperiods, respectively. Directional accuracy remains moderate in the first two periods at 59.76\% and 60.41\%, but declines to 45.10\% in the final period, suggesting a weakening ability to predict return directions despite strong volatility coverage. This is further supported by Figure~\ref{fig:forecast1}-~\ref{fig:forecast2}, where the forecast tracks realized returns closely and the confidence band captures most observed fluctuations.

While the analysis so far has focused on univariate dynamics, transaction activity, prices, and returns, international markets are inherently interconnected. This motivates a transition to a multivariate perspective, where the next sections examine cross country interactions through transaction values and the network structure underlying these relationships.

\section{Inter Country Trading Dynamics}
\begin{figure}[ht]
    \centering \includegraphics[width=0.9\linewidth]{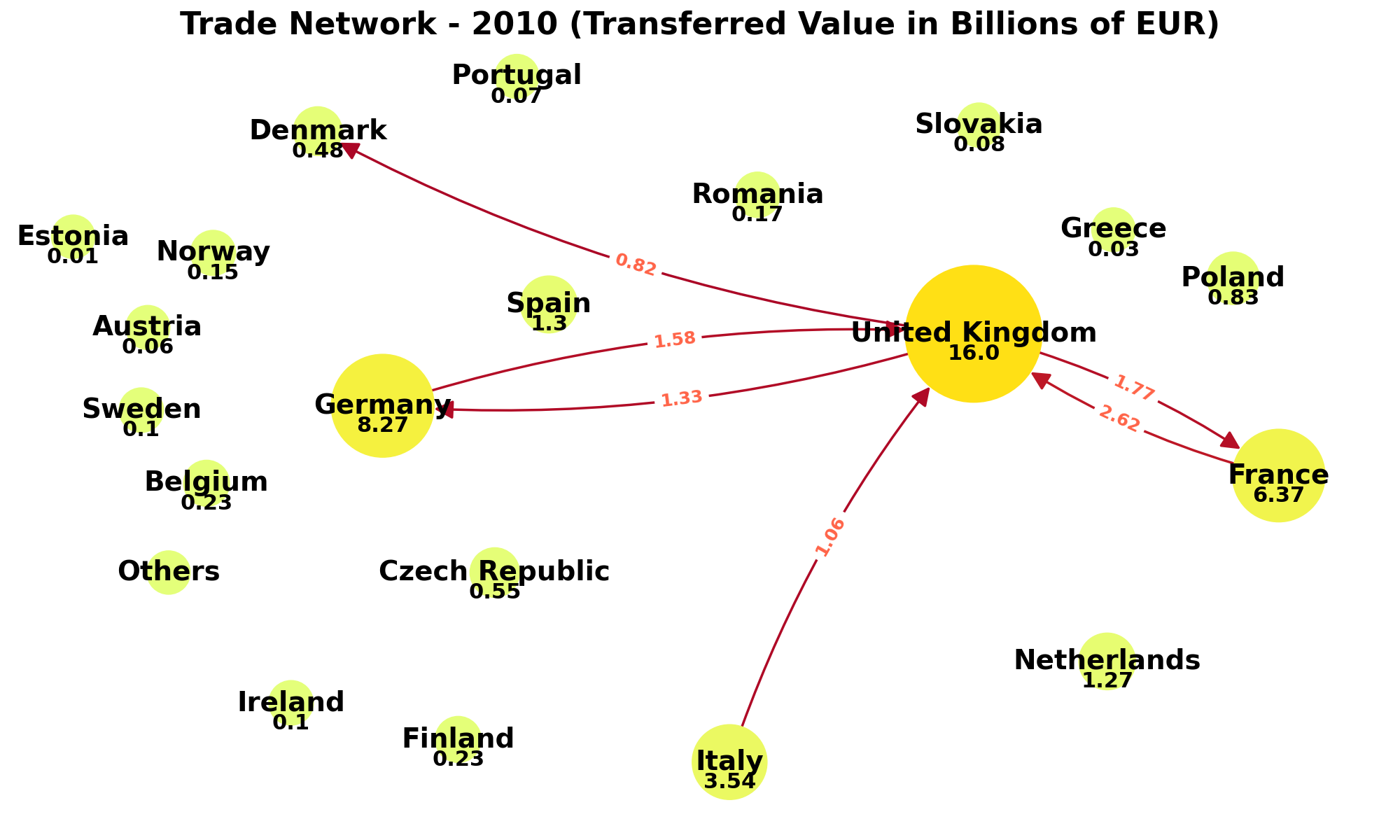}
    \caption{ }
    \label{fig:2010}
\end{figure}

\begin{figure}[ht]
    \centering \includegraphics[width=0.9\linewidth]{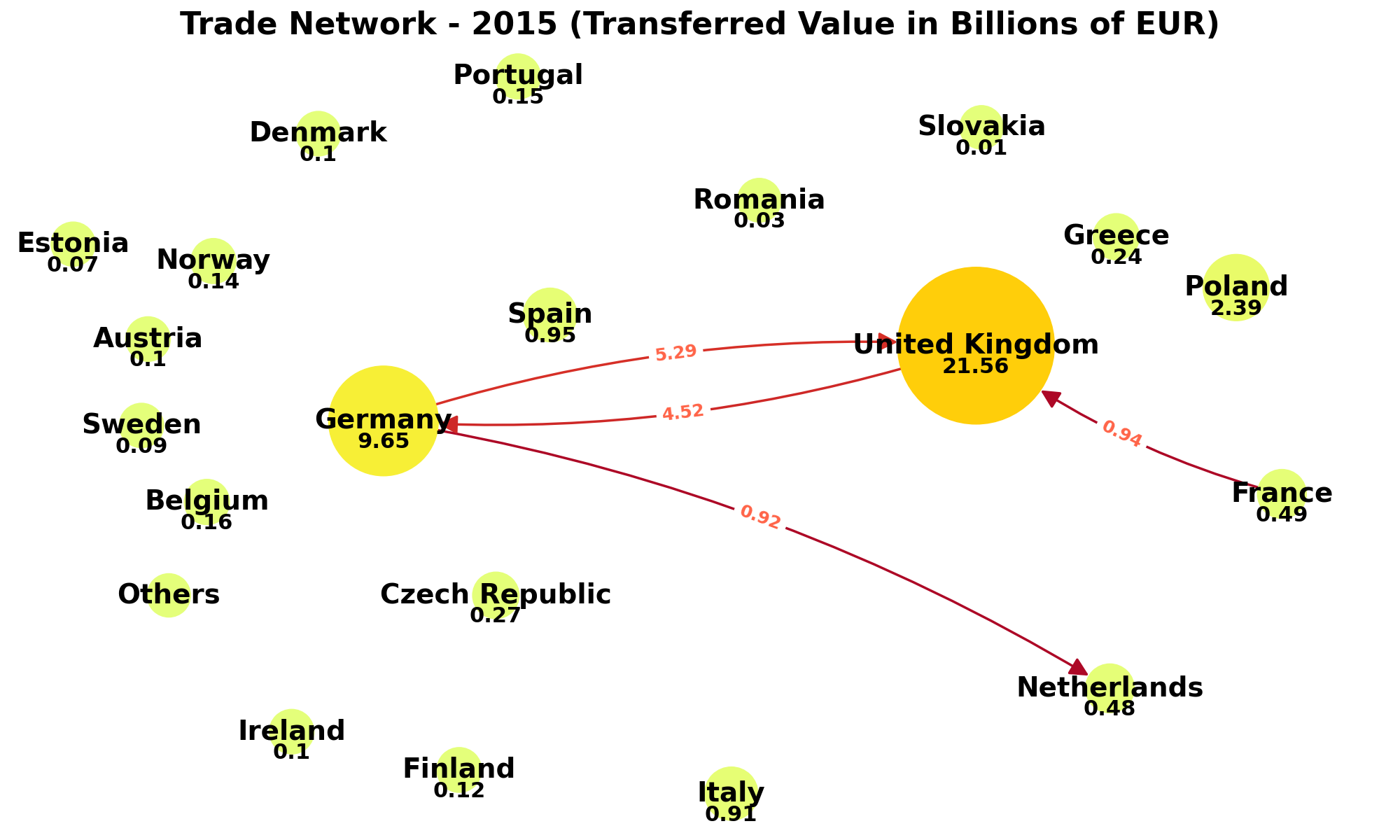}
    \caption{ }
    \label{fig:2015}
\end{figure}

\begin{figure}[H]
    \centering \includegraphics[width=0.9\linewidth]{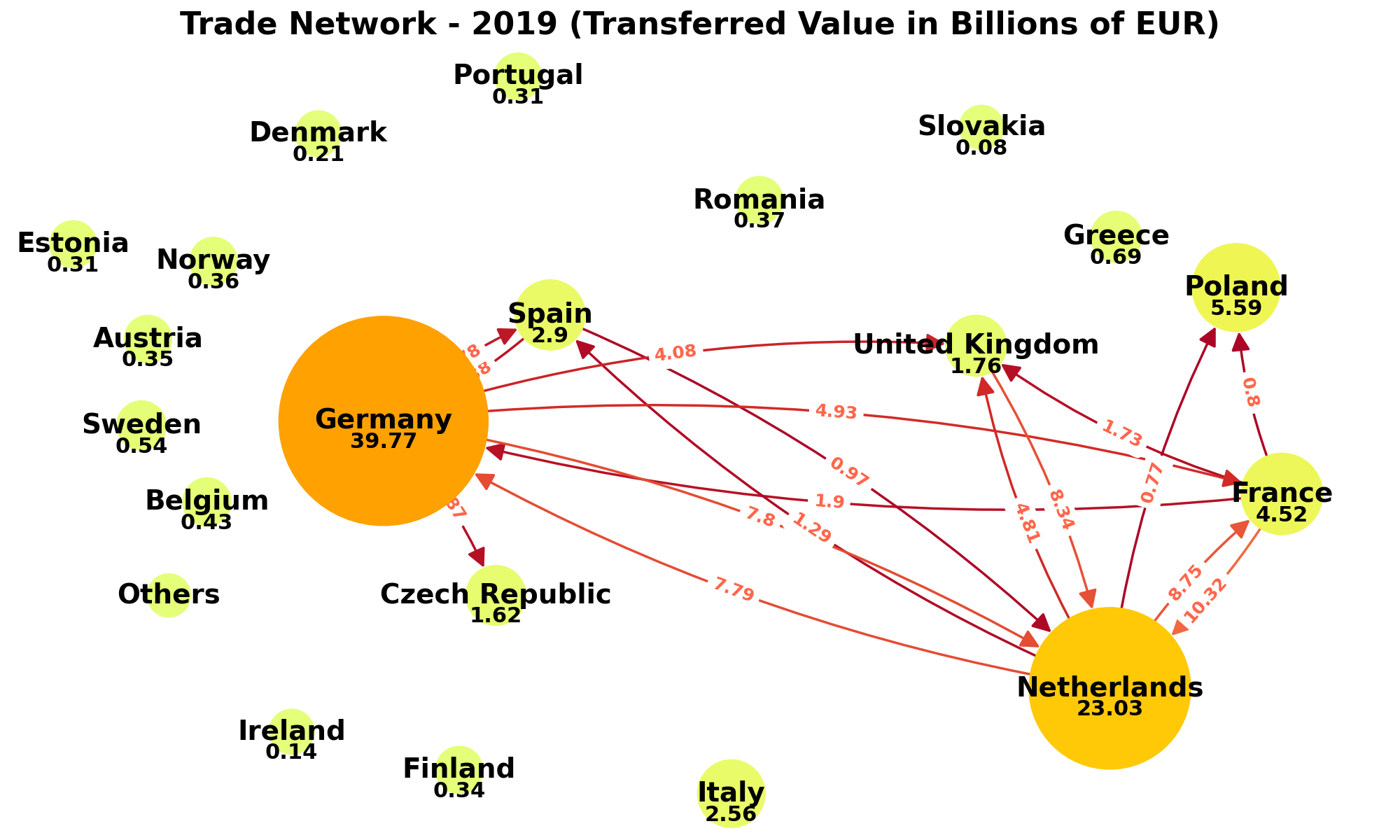}
    \caption{ }
    \label{fig:2019}
\end{figure}

This section examines the evolution of cross-country transaction patterns through the interaction of Operator and Person Holding Accounts (OHA–PHA), which characterize core market linkages and structural shifts. The annual average transaction value (EUR) is calculated for each participating registry. Figures~\ref{fig:2010}-\ref{fig:2019} present the resulting transaction networks for 2010, 2015, and 2019. These years were selected as representatives from the distinct periods previously identified in the price dynamics analysis.

In the network diagrams, nodes signify domestic transactions within a country, and directed edges represent cross-border transfers from a source registry to an acquiring one. A default size is used for all nodes; however, the diameter of a node increases proportionally with its internal transaction value once a specific threshold is exceeded. To maintain visual clarity, only external links surpassing a predefined value are illustrated. Both the edge threshold and the node scaling are applied uniformly across all years to ensure consistent comparability.

The network visualization highlights several notable patterns in the evolution of market participation over time. Prominently, only a handful of countries, roughly four to six-are actively engaged in high-value internal transactions, underscoring a degree of inefficiency and concentration in participation within the system. This limited engagement suggests that the trading activity is driven primarily by a small core of influential countries, while many others remain relatively inactive or peripheral.

A particularly striking observation is the decline in the United Kingdom’s node size in 2019 compared to 2010 and 2015. This reduction likely reflects the uncertainty surrounding Brexit, which diminished the UK's integration and centrality within the EU ETS network during the transition period. In contrast, the 2019 network exhibits higher overall density and volatility, a structural shift that can be plausibly associated with the introduction of the Market Stability Reserve \cite{MSR}.

The MSR’s mechanism to absorb surplus allowances appears to have tightened the market, leading to stronger interconnections and more dynamic trading behavior among participants. Moreover, Germany and the Netherlands emerge as increasingly prominent nodes in 2019. Together, these trends point to a more concentrated yet more active network, shaped by both regulatory reforms and evolving national roles within the carbon market.

Since some countries exhibit particularly large internal as well as external transaction values, it is essential to quantify the \textbf{relative importance of each node} in the directed trade networks over time. To this end, we employ \textbf{Eigenvector Centrality} \cite{network}, \cite{network2}, \cite{network3} to capture not only the direct connectivity of a node but also the influence it derives from its neighbors. 

For a directed, weighted network with adjacency matrix $\mathbf{A}$, where both edges and nodes carry intrinsic weights, we define the \textbf{Eigenvector Centrality} $x_i$ of node $i$ as the solution to the eigenvalue problem:
$$
\mathbf{A} \mathbf{x} = \lambda \mathbf{x},
$$
where $\mathbf{x} = (x_1, x_2, \dots, x_n)^\top$ is the principal eigenvector corresponding to the largest eigenvalue $\lambda$. In our formulation, the weighted adjacency matrix $\mathbf{A}$ incorporates both external trade connections and internal trade capacity:
$$
A_{ij} = 
\begin{cases}
\dfrac{w_{ij}}{\max\{w_{kl}\}}, & \text{if } i \neq j \text{ (external trade)}, \\[10pt]
\dfrac{w_ii}{\max\{w_{kl}\}}, & \text{if } i = j \text{ (self-trade as self-loop)},
\end{cases}
$$
where $w_{ij}$ represents the trade value from country $i$ to country $j$, $w_ii$ denotes the self-trade value of country $i$, and the normalization ensures all weights lie in $[0,1]$. The maximum is taken over all edge weights (including self-loops) to maintain consistent scaling.

The centrality of node $i$ can thus be expressed as:
$$
x_i = \frac{1}{\lambda} \sum_{j=1}^{n} A_{ij} \, x_j = \frac{1}{\lambda} \left( A_{ii} \, x_i + \sum_{j \neq i} A_{ij} \, x_j \right),
$$
where the first term $A_{ii} \, x_i$ captures the contribution of self-trade (internal capacity), reflecting a country's intrinsic importance through its domestic transactions, and the second term $\sum_{j \neq i} A_{ij} \, x_j$ reflects the influence derived from incoming connections, whereby a node gains importance by receiving trade from other highly central nodes. The centrality values $\mathbf{x}$ are estimated by solving the eigenvector problem using power iteration or other numerical eigenvalue methods, specifically computing the principal eigenvector corresponding to the dominant eigenvalue $\lambda$ of the adjacency matrix $\mathbf{A}$. 

\begin{figure}[h]
    \centering
    \includegraphics[width=0.7\linewidth]{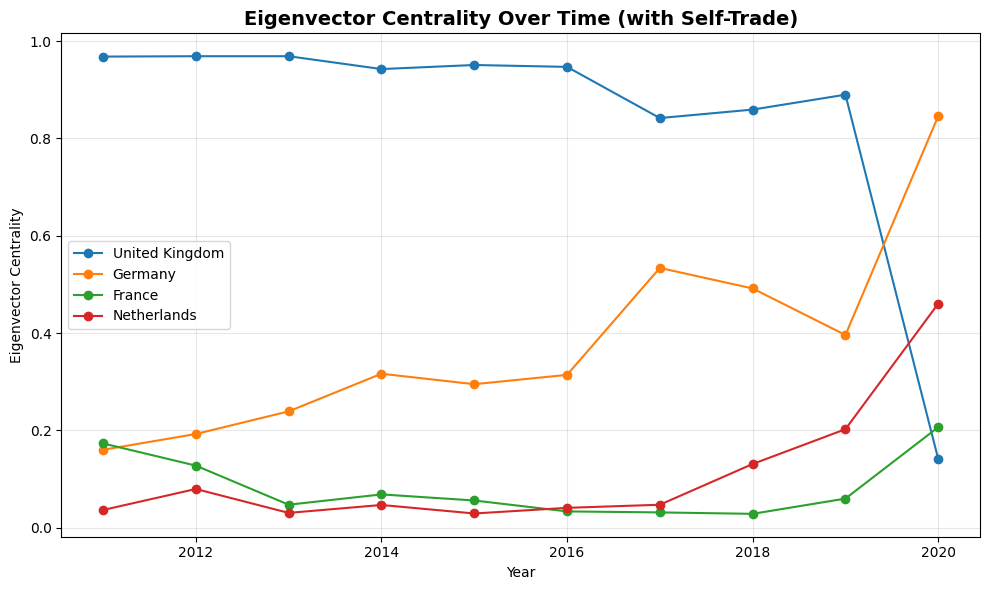}
    \caption{}
    \label{fig:central}
\end{figure}

Since eigenvectors are determined only up to a scalar multiple, the resulting vector $\mathbf{x}$ is typically normalized. To interpret the \textbf{relative importance} of each country, we compute the proportion of total centrality as:
\begin{equation}
p_i = x_i^2,
\end{equation}
where $p_i \in [0,1]$ represents the squared normalized centrality, providing a probabilistic interpretation of node is contribution to the overall network structure. This normalization ensures that $\sum_{i=1}^{n} p_i = 1$, allowing direct comparison of relative influence across countries and over time. Eigenvector centrality therefore provides a nuanced measure of prominence in the network, reflecting both direct transactional activity and the broader position within the network’s structure.

Figure~\ref{fig:central} presents the temporal evolution of Eigenvector Centrality for the leading countries in the trade network, illustrating how the relative influence of each participant evolves over time. The United Kingdom maintained a dominant position for most of the period from 2011 to 2018, contributing around 90\% of the total centrality among the top trading countries. This indicates that the UK was highly connected and linked to other influential countries, giving it a central role in the network. 

In 2020, however, the UK’s centrality dropped sharply, likely reflecting the economic and regulatory uncertainties associated with the Brexit protocol, which disrupted trade patterns. UK left the EU in 2021, it launched its own fully independent carbon market, the \textbf{UK Emissions Trading Scheme} (UK ETS) \cite{ukets}. By contrast, Germany, Netherlands showed a steady increase in centrality over the period, reflecting their growing influence in the trade network through stronger connections with other central participants.

These trends highlight how structural changes in the network, driven by policy or economic events, can alter the relative influence of major trading countries over time. 

\section{Demand and Supply Analysis}
Motivated by the preceding section, we now narrow our analysis to the top three countries, France, Germany, and the United Kingdom, and model the volume of allowances exchanged among them.\footnote{Limiting the analysis to three countries (i.e., a $3 \times 3$ network) ensures tractability; including additional participants would substantially increase complexity.} We adopt a simple \textbf{Price-Quantity model}:
\[
Q = a \times P^b \quad \implies \quad \log Q = \log a + b \times \log P,
\]
where the parameter $b$ can be interpreted as the \textbf{price elasticity of demand} ($E_p$) \cite{varian}. This measures how much the quantity demanded of a good changes in response to a change in its price, holding other factors constant:
\[
b = \frac{\%\ \text{change in quantity}}{\%\ \text{change in price}} = \frac{dQ}{dP}\frac{P}{Q}.
\]

From a market efficiency perspective, these two cases have different implications:

\begin{itemize}
    \item If $|E_p| > 1$: \textbf{Elastic demand} means prices serve as strong signals. Small shifts in allowance prices induce large changes in traded quantities, promoting responsiveness and innovation but also potentially increasing volatility. A positive value of $E_p$ in this case could reflect speculative or compliance-driven trading, where rising prices are accompanied by higher trading volumes, while a negative value indicates efficient, demand-driven adjustment.
    
    \item If $|E_p| < 1$: \textbf{Inelastic demand} implies greater stability and predictability, but weakens the incentive effect, since higher prices do not translate into substantial reductions in traded quantities. 
\end{itemize}

\begin{figure}[H]
    \centering
    \includegraphics[height=0.95\textheight]{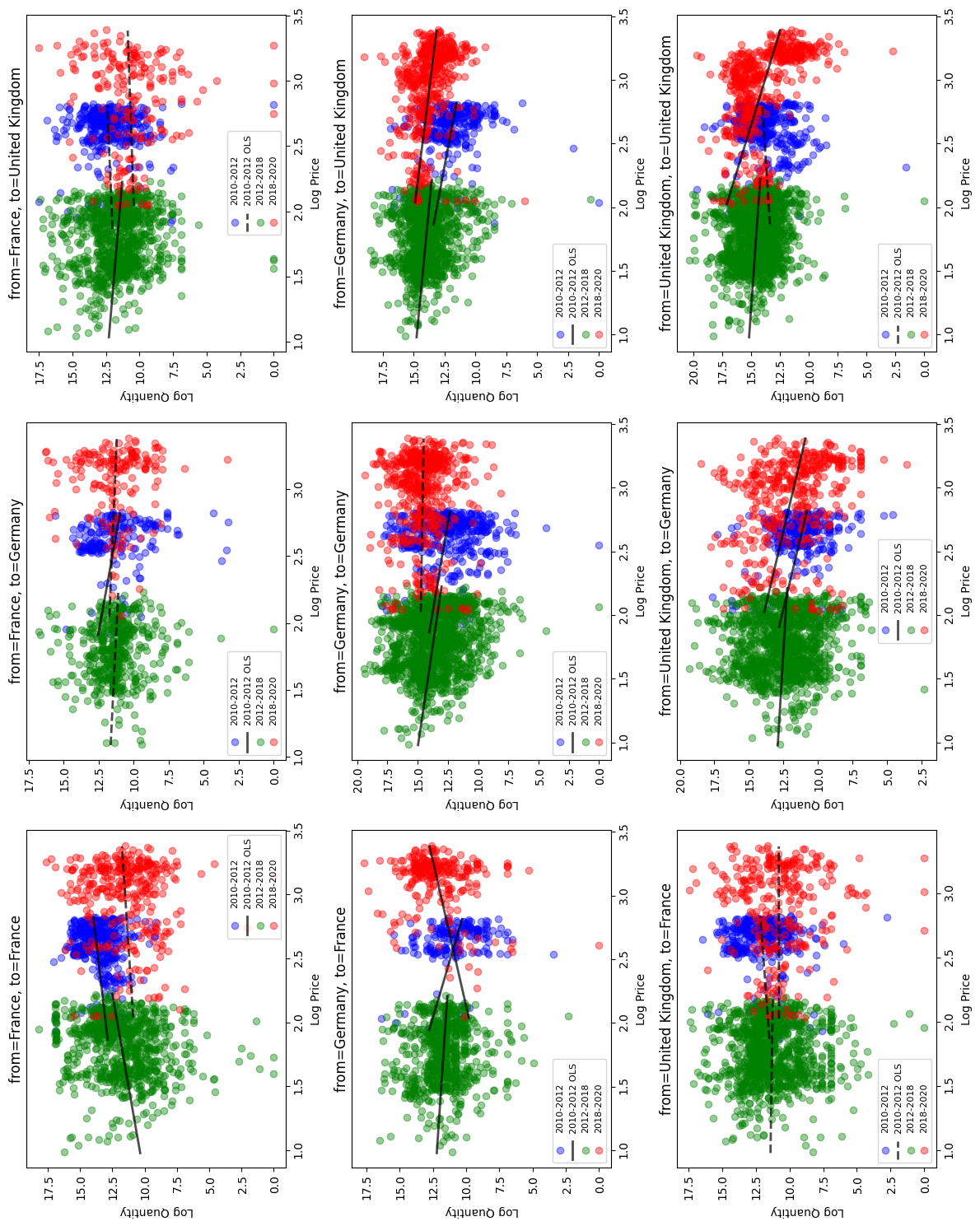}
    \caption{: Transactions of allowances among the top three countries (France, Germany, and the United Kingdom). OLS fits.}
    \label{fig:elasticity}
\end{figure}

\begin{figure}[H]
    \centering
    \includegraphics[width=0.8\linewidth]{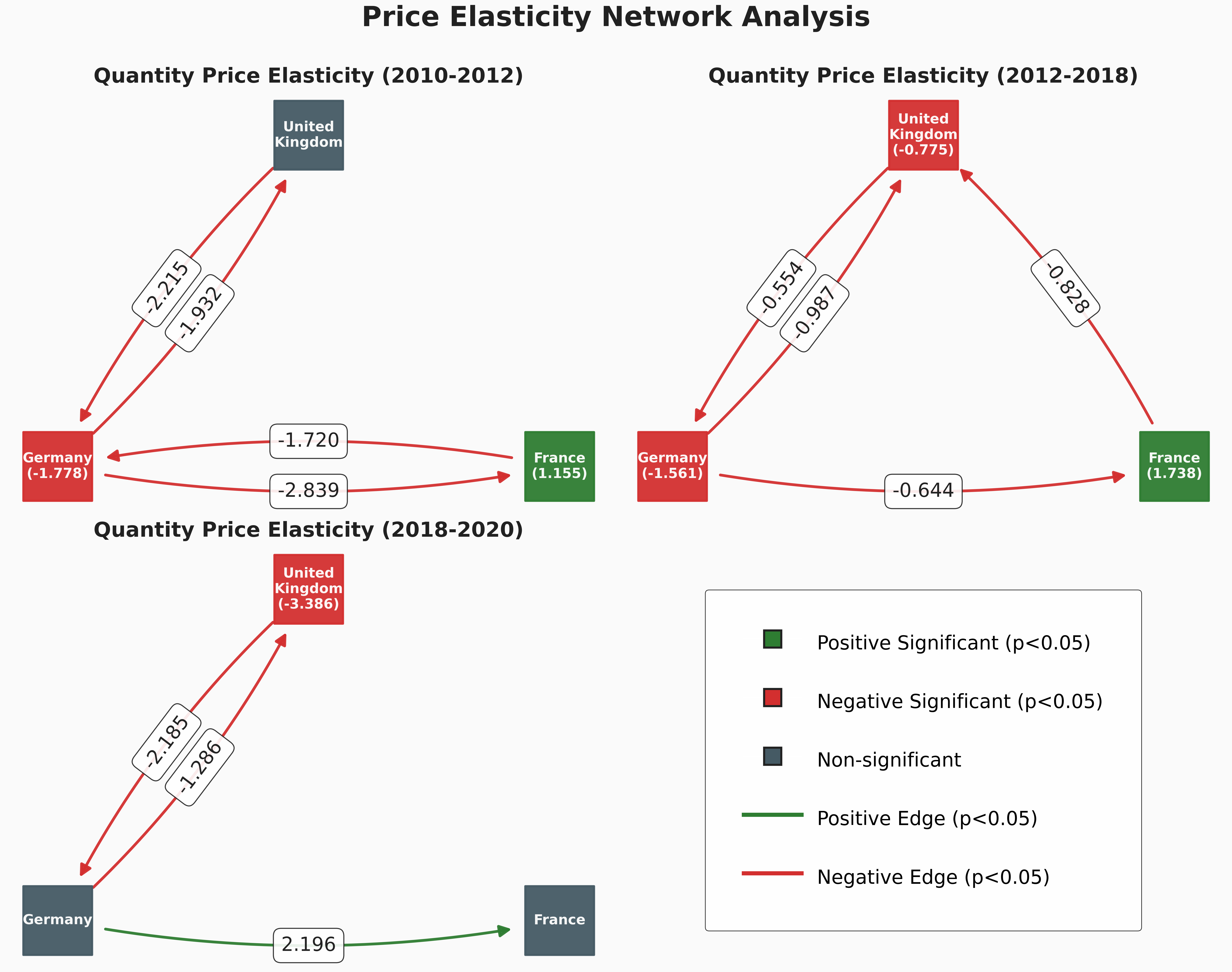}
    \caption{ }
    \label{fig:elasticity2}
\end{figure}

Appendix Table~\ref{tab:model} presents the detailed model summaries. The results indicate that both OLS and LAD estimations yield comparable elasticity patterns across country pairs and periods. In the figure~\ref{fig:elasticity}, we examine the daily log of transaction quantities against the log of prices. The price variable is clustered by periods as defined earlier, represented with three distinct colors. The \textbf{scatter plot} displays the relationship between $\log(P)$ and $\log(Q)$, with points color-coded by \textbf{time period} to highlight temporal variation. The \textbf{black lines} represent the \textbf{Ordinary Least Square} (OLS) regression fits of $\log(Q)$ on $\log(P)$ for each period. \textbf{Solid lines} correspond to \textbf{statistically significant} fits (low p-values), while \textbf{dotted lines} indicate \textbf{insignificant} relationships.
\newpage
\newpage
We summarize the scatterplot-based price elasticities in a graphical representation in Figure~\ref{fig:elasticity2}. In these graphs, the \textbf{nodes} represent the internal price elasticity of transactions within each country, while the \textbf{edges} capture the external price elasticity of transactions between countries. The insignificant price elasticity estimates for external transactions are not shown in the diagram as edges. Three separate graphs are plotted, corresponding to the three time periods considered.  

The analysis reveals distinct phases in \textbf{market responsiveness} over time. During \textbf{2010-2012}, the edges exhibit \textbf{elastic relationships}, indicating that the market was highly responsive even at relatively low price levels. This suggests that firms had access to \textbf{low-cost abatement options} or \textbf{substitutes}, enabling rapid adjustments in trading behavior in response to price changes. In contrast, between \textbf{2012 and 2018}, nearly all edges appear \textbf{inelastic}, implying reduced responsiveness to price signals, likely reflecting inefficiencies arising from an \textbf{oversupply of allowances} \cite{allowance surplus}, which weakened price sensitivity as discussed earlier. In the \textbf{2018–2020} period, both \textbf{high price volatility} and \textbf{high elasticity} emerge across most edges, signaling a phase of \textbf{market instability} \cite{MSR, ukets}. Participants reacted strongly to price movements, driven by \textbf{allowance scarcity}, \textbf{tighter regulations}, and \textbf{policy uncertainty}, all of which amplified trading responses.

Notably, the \textbf{UK to France} and \textbf{France to Germany} connections do not appear statistically significant (estimated elasticity $\approx 0$) in almost all the period, implying that traded quantities with France were largely unresponsive to price changes, potentially due to \textbf{limited bilateral trading activity} or \textbf{structural barriers} to cross-border allowance flows. Even the intra-country transactions within \textbf{France} showed a \textbf{high positive price elasticity} in initial two periods, suggesting possible \textbf{market inefficiencies}. As prices rose, firms traded more actively, likely driven by \textbf{speculative motives} rather than genuine compliance needs.

We observe a \textbf{non-monotonic evolution} in market behavior: an initial elastic phase driven by abundant low-cost abatement opportunities, followed by an inelastic phase marked by oversupply and weak price signals, and culminating in a highly elastic yet volatile regime shaped by regulatory reforms and supply constraints. These patterns highlight the \textbf{critical role of policy design},particularly in managing allowance supply and harmonizing cross-border mechanisms, to sustain efficient and stable carbon markets. Policymakers must balance \textbf{allowance scarcity} to incentivize abatement with \textbf{market stability} to preserve investor confidence and long-term de-carbonization goals.

\section{Conclusion}
The EU Emissions Trading System (EU ETS), launched in 2005, stands as the world's first and largest international carbon market, covering approximately 40\% of the EU's greenhouse gas emissions across power generation, manufacturing, and aviation sectors. Since its inception, the EU ETS has successfully contributed to a \textbf{reduction of over 35\% in emissions} from covered sectors while maintaining economic competitiveness, demonstrating that market-based mechanisms can effectively balance environmental objectives with economic growth. Through iterative reforms including the introduction of the Market Stability Reserve (MSR) and progressive tightening of the emissions cap the system has evolved into a robust policy instrument that not only incentivizes low-carbon innovation but also serves as a global benchmark for emissions trading schemes worldwide.

Our analysis provides a comprehensive understanding of the dynamics within the European carbon market by examining how market behavior evolved across different policy regimes. We systematically investigate patterns of price movements, transaction dynamics, and inter-country interactions to characterize the structural evolution of the EU ETS.

The evidence from short-term return predictability analysis indicates inefficiencies, with significant autocorrelation in returns suggesting deviations from the efficient market hypothesis. These findings motivated our network-based approach to capturing the international dimension of the market. The network analysis reveals substantial structural shifts in transaction flows over time, characterized by a \textbf{core-periphery structure} in which a small subset of countries, predominantly the United Kingdom, Germany, France, and the Netherlands, occupy central positions as dominant trading hubs, while the majority of participating nations remain at the periphery with limited connectivity. The price elasticity analysis further elucidates heterogeneous trading responses to price fluctuations across distinct policy phases. Both intra- and inter-country trades exhibited markedly different elasticities depending on the regulatory environment, reflecting the interplay between \textbf{speculative activity} and \textbf{compliance-driven motives}. Notably, we also observe \textbf{country-specific heterogeneity} in elasticity patterns for major participants.

Collectively, these findings highlight the \textbf{complex interplay} between policy design, market structure, and trading behavior in the EU ETS, emphasizing the need for coordinated regulatory frameworks that account for national heterogeneity while promoting market integration and efficiency.

Phase 4 (2021–2030) of the EU ETS outlines an optimistic trajectory toward a decarbonized European economy. With a progressively tightening emissions cap, a refined Market Stability Reserve, and an expanded sectoral scope, it enhances price signals and fosters low-carbon investment. If effectively implemented, Phase 4 has the potential to transform the EU ETS into a resilient, innovation-driven market that efficiently drives emission reductions while supporting Europe’s long-term climate ambitions. Our analysis presents a novel perspective on market behavior within the EU ETS, offering new insights into its evolving efficiency and responsiveness. While preliminary, this study opens multiple avenues for further research, including analyses of surrender data, more advanced investigations into return predictability, and a deeper exploration of price elasticity dynamics. Such extensions would enrich the understanding of how market mechanisms can effectively support Europe’s transition toward a low-carbon future.

Link to the codes used in the study: \href{https://github.com/Avirup23/Carbon-Market-Analysis.git}{Github}

\newpage
\section*{Appendix}

\begin{table}[htbp]
\centering
\caption{\textbf{OLS and LAD Regression Results: Log Quantity on Log Price by Country Pair and Period}}
\label{tab:model}
\vspace{2mm}
\resizebox{\textwidth}{!}{%
\renewcommand{\arraystretch}{1.1}
\begin{tabular}{|llc|cccc|cccc|c|}
\toprule
\multirow{2}{*}{\textbf{From}} & \multirow{2}{*}{\textbf{To}} & \multirow{2}{*}{\textbf{Period}} &
\multicolumn{4}{c|}{\textbf{OLS}} & \multicolumn{4}{c|}{\textbf{LAD}} & \multirow{2}{*}{\textbf{N}} \\ 
\cmidrule(lr){4-7} \cmidrule(lr){8-11}
 &  &  & $\hat{\beta}_0$ & $p$-val & $\hat{\beta}_1$ & $p$-val & $\hat{\beta}_0$ & $p$-val & $\hat{\beta}_1$ & $p$-val & \\ 
\midrule
France & France & 2010–2012 & 10.70 & 0.000 & 1.15*** & 0.000 & 8.42 & 0.000 & 2.08*** & 0.000 & 495 \\ 
France & France & 2012–2018 & 8.63 & 0.000 & 1.74*** & 0.000 & 7.87 & 0.000 & 2.22*** & 0.000 & 843 \\ 
France & France & 2018–2020 & 9.55 & 0.000 & 0.65 & 0.131 & 8.08 & 0.000 & 1.09 & 0.058 & 347 \\ 
\midrule
France & Germany & 2010–2012 & 15.84 & 0.000 & -1.72* & 0.036 & 15.30 & 0.000 & -1.35 & 0.144 & 214 \\ 
France & Germany & 2012–2018 & 12.17 & 0.000 & -0.47 & 0.180 & 12.33 & 0.000 & -0.53 & 0.065 & 508 \\ 
France & Germany & 2018–2020 & 12.55 & 0.000 & -0.40 & 0.496 & 12.29 & 0.000 & -0.32 & 0.626 & 182 \\ 
\midrule
France & UK & 2010–2012 & 11.23 & 0.000 & 0.42 & 0.392 & 11.11 & 0.000 & 0.55 & 0.278 & 387 \\ 
France & UK & 2012–2018 & 13.15 & 0.000 & -0.83** & 0.002 & 13.49 & 0.000 & -1.07*** & 0.000 & 972 \\ 
France & UK & 2018–2020 & 9.73 & 0.000 & 0.34 & 0.490 & 10.57 & 0.000 & 0.08 & 0.875 & 201 \\ 
\midrule
Germany & France & 2010–2012 & 18.31 & 0.000 & -2.84*** & 0.001 & 18.53 & 0.000 & -2.88** & 0.009 & 179 \\ 
Germany & France & 2012–2018 & 12.83 & 0.000 & -0.64* & 0.016 & 12.50 & 0.000 & -0.49 & 0.055 & 750 \\ 
Germany & France & 2018–2020 & 5.29 & 0.004 & 2.20*** & 0.000 & 2.64 & 0.154 & 3.11*** & 0.000 & 267 \\ 
\midrule
Germany & Germany & 2010–2012 & 17.37 & 0.000 & -1.78** & 0.002 & 19.67 & 0.000 & -2.65*** & 0.001 & 467 \\ 
Germany & Germany & 2012–2018 & 16.51 & 0.000 & -1.56*** & 0.000 & 16.41 & 0.000 & -1.42*** & 0.000 & 1467 \\ 
Germany & Germany & 2018–2020 & 15.09 & 0.000 & -0.17 & 0.427 & 15.18 & 0.000 & -0.19 & 0.431 & 581 \\ 
\midrule
Germany & UK & 2010–2012 & 16.99 & 0.000 & -1.93*** & 0.001 & 18.50 & 0.000 & -2.45*** & 0.000 & 322 \\ 
Germany & UK & 2012–2018 & 15.73 & 0.000 & -0.99*** & 0.000 & 15.39 & 0.000 & -0.67*** & 0.000 & 1353 \\ 
Germany & UK & 2018–2020 & 17.50 & 0.000 & -1.29*** & 0.000 & 17.77 & 0.000 & -1.37*** & 0.000 & 553 \\ 
\midrule
UK & France & 2010–2012 & 10.16 & 0.000 & 0.71 & 0.184 & 9.06 & 0.000 & 1.11 & 0.136 & 370 \\ 
UK & France & 2012–2018 & 11.54 & 0.000 & -0.14 & 0.627 & 11.79 & 0.000 & -0.14 & 0.646 & 1109 \\ 
UK & France & 2018–2020 & 10.75 & 0.000 & 0.01 & 0.978 & 11.59 & 0.000 & -0.23 & 0.605 & 310 \\ 
\midrule
UK & Germany & 2010–2012 & 17.06 & 0.000 & -2.21*** & 0.000 & 17.42 & 0.000 & -2.35*** & 0.000 & 310 \\ 
UK & Germany & 2012–2018 & 13.46 & 0.000 & -0.55* & 0.036 & 12.60 & 0.000 & -0.11 & 0.719 & 1279 \\ 
UK & Germany & 2018–2020 & 18.33 & 0.000 & -2.19*** & 0.000 & 21.04 & 0.000 & -3.10*** & 0.000 & 432 \\ 
\midrule
UK & UK & 2010–2012 & 12.01 & 0.000 & 0.71 & 0.094 & 10.27 & 0.000 & 1.46*** & 0.000 & 486 \\ 
UK & UK & 2012–2018 & 15.92 & 0.000 & -0.77*** & 0.000 & 15.41 & 0.000 & -0.52* & 0.022 & 1481 \\ 
UK & UK & 2018–2020 & 23.96 & 0.000 & -3.39*** & 0.000 & 25.84 & 0.000 & -3.94*** & 0.000 & 534 \\ 
\bottomrule
\end{tabular}}
\vspace{1mm}
\scriptsize
\textit{Note:} The model is $\ln(\text{Quantity}) = \beta_0 + \beta_1 \ln(\text{Price}) + \epsilon$. OLS = ordinary least squares; LAD = least absolute deviation (median regression). $\hat{\beta}_1$ represents price elasticity. Significance: * $p<0.05$, ** $p<0.01$, *** $p<0.001$. \textit{N} = number of observations.
\end{table}


\begin{table}[h!]
\centering
\caption{Rolling Forecast Evaluation Across Periods}
\label{tab:evaluation}
\begin{tabular}{lcccc}
\toprule
\textbf{Period} & \textbf{Model} & \textbf{Directional Accuracy} & \textbf{Hit Rate ($\pm 2\sigma$)} & \textbf{Observations} \\
\midrule
2010--2012 & ARMA(1,0)-GARCH(1,1) & 59.76\% & 93.90\% & 82 \\
2012--2018 & ARMA(3,0)-GARCH(1,1) & 60.41\% & 83.96\% & 293 \\
2018--2020 & ARMA(1,0)-GARCH(1,1) & 45.10\% & 93.14\% & 102 \\
\bottomrule
\end{tabular}
\end{table}

\begin{figure}[H]
    \centering    \includegraphics[width=0.90\linewidth]{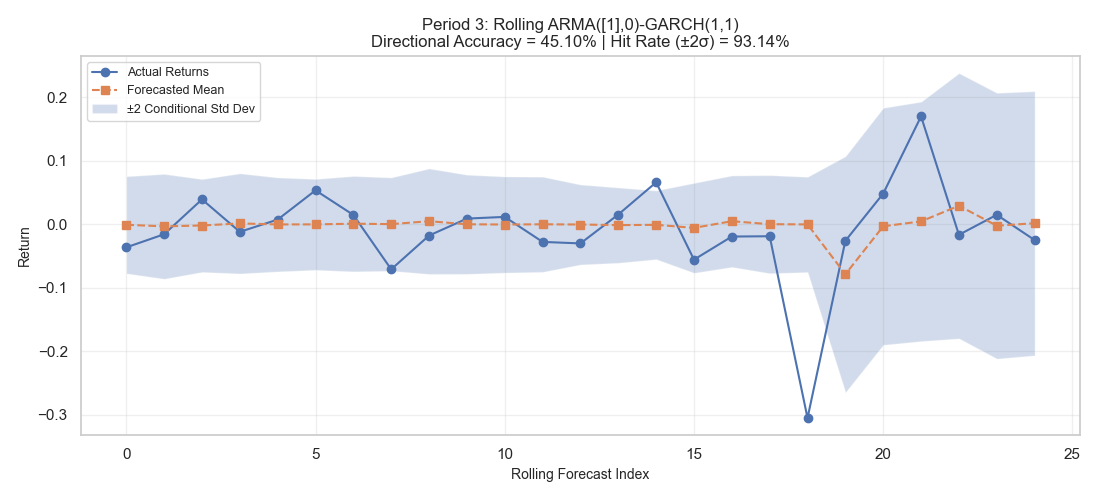}
    \caption{ }
    \label{fig:forecast3}
\end{figure}

\newpage
\begin{thebibliography}{99}

\bibitem{database} 
Abrell, J. (2024). \textit{Database for the European Union Transaction Log}. Available at \url{https://www.euets.info}.

\bibitem{inefficiency} 
Borri, N., Liu, Y., Tsyvinski, A., \& Wu, X. (2024). \textit{Inefficiencies of Carbon Trading Markets}. \url{https://arxiv.org/abs/2408.06497}.

\bibitem{inefficiency2} 
Hintermann, B., Peterson, S., \& Rickels, W. (2016). \textit{Price and Market Behavior in Phase II of the EU ETS: A Review of the Literature}. Review of Environmental Economics and Policy, 10(1), 108–128.

\bibitem{inefficiency3} 
Ibikunle, G., Gregoriou, A., Hoepner, A. G., \& Rhodes, M. (2016). \textit{Liquidity and Market Efficiency in the World’s Largest Carbon Market}. British Accounting Review, 48(4), 431–447.

\bibitem{zhikai2024volatility} 
Zhang, Z., Li, S., \& Zhang, W. (2024). \textit{The Predictability of Carbon Futures Volatility: New Evidence from the Spillovers of Fossil Energy Futures Returns}. Journal of Futures Markets, 44(3), 512–530.

\bibitem{zhao2024forecasting} 
Zhao, Y., Gong, X., Zhang, W., \& Xu, W. (2024). \textit{Forecasting Carbon Futures Returns Using Feature Selection and Markov Chain with Sample Distribution}. Energy Economics, 140, 108079.

\bibitem{zhang2022liquidity} 
Zhang, J., \& Han, W. (2022). \textit{Carbon Emission Trading and Equity Markets in China: How Liquidity is Impacting Carbon Returns?} Economic Research-Ekonomska Istraživanja, 35(1), 402–420.

\bibitem{kim2021liquidity} 
Kim, J., \& Park, K. (2021). \textit{Improving Liquidity in Emission Trading Schemes}. Journal of Futures Markets, 41(9), 1397–1411.

\bibitem{EUETS.INFO}
EUETS.INFO \url{https://www.euets.info/download}.

\bibitem{ICAP}
ICAP. \url{https://icapcarbonaction.com/en/ets-prices}.

\bibitem{allowance surplus}
European Commission. (2025). \textit{Surplus of Allowances}. Retrieved from \url{https://climate.ec.europa.eu/eu-action/carbon-markets/eu-emissions-trading-system-eu-ets/market-stability-reserve_en#:~:text=In%202013%2C%20the,381%20million%20allowances}

\bibitem{MSR}
European Commission. (2025). \textit{Market Stability Reserve}. Retrieved from \url{https://climate.ec.europa.eu/eu-action/carbon-markets/eu-emissions-trading-system-eu-ets/market-stability-reserve_en#:~:text=The%20Market%20Stability%20Reserve%20(MSR)%20was,2018%20and%20began%20operating%20in%202019.}

\bibitem{Market efficiency}
Döme, B. (2005). \textit{Efficiency and Return Predictability: What can we learn from stock market data?} C.U. Budapest. May 2, 2005.

\bibitem{ukets}
Gov UK. \textit{UK ETS} \url{https://www.gov.uk/government/publications/uk-emissions-trading-scheme-uk-ets-policy-overview/uk-emissions-trading-scheme-uk-ets-a-policy-overview?utm_source=chatgpt.com#what-is-the-uk-ets:~:text=The%20UK%2C%20Scottish,1%20January%202021.}

\bibitem{network}
Horvath, S. (2011). \textit{Weighted network analysis: Applications in genomics and systems biology.} Springer.

\bibitem{network2}
Wasserman, S., \& Faust, K. (1994). \textit{Social network analysis: Methods and applications.} Cambridge University Press. \url{https://doi.org/10.1017/CBO9780511815478}

\bibitem{network3}
Bloch, F., Jackson, M. O., \& Tebaldi, P. (2021). \textit{Centrality measures in networks}. \url{https://arxiv.org/pdf/1608.05845}

\bibitem{varian}
Varian, H. R. (2014). \textit{Intermediate Microeconomics: A Modern Approach (9th ed.)}. W.W. Norton \& Company.
\end{thebibliography}
\end{document}